\documentclass{appolb}
\usepackage{graphicx}
\usepackage{cite}
\usepackage{hyperref}
\usepackage{amsmath}
\usepackage{epstopdf}%
\usepackage{amsfonts}%
\usepackage{amssymb}
\usepackage{tikz}

\DeclareMathOperator{\disc}{Disc}

\begin{document}

\title{The scalar--isovector sector in the extended Linear Sigma Model%
\thanks{Presented at the Workshop ``Excited QCD 2014'', Bjelasnica Mountain, Sarajevo, Bosnia--Herzegovina, February 2--8, 2014.}%
}
\author{T. Wolkanowski and F. Giacosa
\address{Institut f\"{u}r Theoretische Physik, Goethe-Universit\"at Frankfurt am Main Max-von-Laue-Str. 1, 60438 Frankfurt am Main, Germany}
}
\maketitle

\begin{abstract}
We study basic properties of scalar hadronic resonances within the so-called extended Linear Sigma Model (eLSM), which is an effective model of QCD based on chiral symmetry and dilatation invariance. In particular, we focus on the mass and decay width of the scalar isovector state $a_{0}(1450)$ and perform a numerical study of the propagator pole(s) on the unphysical Riemann sheets. In this work, this meson is understood as a seed state explicitly included in the eLSM. Our results show that the inclusion of hadronic loops does not modify much the previously obtained tree-level results. Moreover, the $a_{0}(980)$ cannot be found as a propagator pole generated by hadronic loop contributions.
\end{abstract}

  
\section{Introduction}
It is nowadays recognized that the scalar sector of hadronic particles is not well described by the ordinary $q\bar{q}$ picture based on a simple representation of $SU(3)$ flavour symmetry \cite{amslerrev}. One very simple reason is the mere fact that the number of physical resonances is much larger than the number of states that can be constructed within a $q\bar{q}$ picture. In particular, in the scalar--isovector sector it is possible to build up only one such state, though two isotriplets are definitely established, the resonances $a_{0}(980)$ and $a_{0}(1450)$ \cite{beringer,baker_bugg}. The question is not only which model or interpretation should be used to handle some of the scalars accurately, but to get mostly {\em all} of them right ({\em i.e.}, the masses and decay widths).

If we write down a Lagrangian with only a single scalar state (as, for instance, the resonance $a_{0}(1450)$) we can assign a free propagator to the mesonic state, which becomes dressed if the coupling to a decay channel is non-zero. One sometimes speaks of mesonic clouds in this context \cite{salam,achasov,giacosaSpectral,e38,coito}. The mass and width of the resonance is then determined by the position of the complex pole of the full interacting propagator on the appropriate unphysical Riemann sheet -- a procedure first proposed by Peierls \cite{peierls} a long time ago, making the quantum theoretical treatment of unstable particles to become an object of much interest (see Refs. \cite{levypoles,aramaki,landshoff} for only some articles published after Peierl's work).

Soon after T\"ornqvist studied the scalar sector by including hadronic loop contributions in the 1990s \cite{tornqvist}, Boglione and Pennington suggested a new kind of mechanism of {\em dynamical generation} of states \cite{pennington} (this term is not used consistently by all authors working in the field, see Ref. \cite{giacosaDynamical} for an overview). According to them, because of this strong coupling to intermediate states the scalar sector not only escapes from the general approach of the naive quark model, but additional resonances with the same quantum numbers can be generated: In the isovector sector ($I=1$) only one seed state, roughly corresponding to $a_{0}(1450)$, is present; then, due to quantum corrections, a second pole, corresponding to the resonance $a_{0}(980)$ is dynamically generated (see also Ref. \cite{eef,dullemond}). Note, alternative interpretations exist: $a_{0}(980)$ is often regarded as a kaonic bound state \cite{lohse,oset}, since it lies just below the $K\bar{K}$ threshold, or as a tetraquark state \cite{tq,tqmix}.

The extended Linear Sigma Model (eLSM) is an effective model of QCD with (pseudo)scalar as well as (axial-)vector states based on chiral symmetry and dilatation invariance \cite{eLSM1,eLSM2}. In this model, the scalar--isovector state is identified with the resonance $a_{0}(1450).$ In this paper, we use the parameters of Ref. \cite{eLSM2} in order to calculate the propagator of $a_{0}(1450)$: in this way we can test the effect of loops on this broad scalar state. Then, we focus on the region below the $K\bar{K}$ threshold and try to find out whether $a_{0}(980)$ emerges as a companion pole in the propagator. The answer is negative.

\section{Used model and method}
The mesonic part of the eLSM includes (pseudo)scalar and (axial-)vector mesons both in the non-strange and strange sectors, where all of them are assigned as $q\bar{q}$ states. Furthermore, all needed symmetries and symmetry breaking terms are present: on the one hand the effective Lagrangian possesses a chiral $U(3)_{L}\times U(3)_{R}$ symmetry and dilatation invariance, while on the other hand explicit (due to non-vanishing quark masses) as well as spontaneous symmetry breaking (due to a non-vanishing chiral condensate $\langle q\bar{q}\rangle\neq0$) and the $U(1)_{A}$ chiral anomaly are taken into account. Concerning dilatation invariance, this symmetry is explicitly broken because of the trace anomaly by including a dilaton/glueball field.

It was shown in a variety of publications (see {\em e.g.} Refs. \cite{eLSM1,eLSM2} and references therein) that the overall phenomenology of the model is good, in other words it is possible to describe many different hadrons within a unified model. The best fit of the model parameters to the available data needs the scalar resonance(s) with isospin $I=1$ (and $I=1/2$) to lie above 1 GeV and to be quark--antiquark states. The scalar isotriplet was consequently identified with $a_{0}(1450)$ with a fitted mass of $1363$ MeV. This is the bare mass as it appears in the Lagrangian; it will be denoted as $M_{0}$. For our purpose it is sufficient to give only the relevant interaction part of the Lagrangian for the neutral state $a_{0}^{0}$
\begin{equation}
\mathcal{L}_{\text{int}} = Aa_{0}^{0}\eta\pi^{0}+Ba_{0}^{0}\eta^{\prime}\pi^{0}+Ca_{0}^{0}(K^{0}\bar{K}^{0}-K^{-}K^{+}) \ ,
\end{equation}
where $\pi^{0},\eta,\eta^{\prime},K$ are the pseudoscalar mesons, and the constants $A,B,C$ are combinations of the coupling constants and masses taken from Ref. \cite{eLSM2}. They are constructed in such a way that the decay amplitude for each channel, $-i\mathcal{M}_{ij}$, is momentum independent. The optical theorem for Feynman diagrams can then be applied to compute the imaginary part of the corresponding self-energy loop $\Pi_{ij}(s)$, regularized by a Gaussian 3d-cutoff function with cutoff scale $\Lambda=0.85$ GeV \cite{close}:
\begin{equation}
\int\text{d}\Gamma \ |\text{--}i\mathcal{M}_{ij}|^{2} = \sqrt{s} \ \Gamma_{ij}^{\text{tree}}(s) = -\operatorname{Im}\Pi_{ij}(s) \ , \ \ \Gamma_{ij}^{\text{tree}} = \frac{|\textbf{k}|}{8\pi M_{0}^{2}}|\text{--}i\mathcal{M}_{ij}|^{2} \ ,
\end{equation}
where $\textbf{k}$ is the three-momentum of one of the emitted particles in the decay of the $a_{0}^{0}$ in its rest frame. The real part is obtained by the dispersion relation
\begin{equation}
\operatorname{Re}\Pi_{ij}(s) = \frac{1}{\pi} -\!\!\!\!\!\!\!\int\text{d}s^{\prime} \ \frac{-\operatorname{Im}\Pi_{ij}(s^{\prime})}{s-s^{\prime}} \ .
\end{equation}
After that, the self-energy is analytically continued to complex values, $s\rightarrow z$, while the continuation into the appropriate unphysical Riemann sheet(s) can be done by exploiting the idea of a Riemann surface and adding the discontinuities for each channel
\begin{equation}
\Pi_{ij}^{c}(z) = \Pi_{ij}(z)+\disc\Pi_{ij}(z) \ , \ \ \disc\Pi_{ij}(s) = 2i\lim_{\epsilon \to 0^{+}}\operatorname{Im}\Pi_{ij}(s+i\epsilon) \ , \label{equation_contself}
\end{equation}
where the superscript $c$ indicates the continued function on the next sheet. Note that in this paper the appropriate sheet is taken to be the one closest to the physical region.

In the so-called one-loop approximation the hadronic loop contributions appear in the inverse expression of the full interacting propagator of the scalar resonance $a_{0}^{0}$ after Dyson resummation
\begin{equation}
\Delta^{-1}(s)=s-M_{0}^{2}-\Pi(s) \ ,
\end{equation}
where here $\Pi(s)$ is the sum of all contributing channels. The propagator on the unphysical sheet(s) is obtained by replacing the suitable self-energy functions according to the first expression of Eq. (\ref{equation_contself}).

\section{Results and discussions}
From a phenomenological point of view it is natural to think of the bare mass $M_{0}$ as being highly influenced by the strong coupling to intermediate hadronic states and the creation of a mesonic cloud, respectively. For instance, we have shown in Ref. \cite{e38} in the case of only one channel that such kind of mechanism usually causes the physical mass (as the real part of the propagator pole) to be smaller than $M_{0}$ (if the coupling constant is not to large). However, this changes now: the mass is larger due to the interplay of various decay channels.
\begin{figure}[b]
\centerline{
\begin{tikzpicture}
\node (L085trajectory) {\centerline{\includegraphics[scale=0.85]{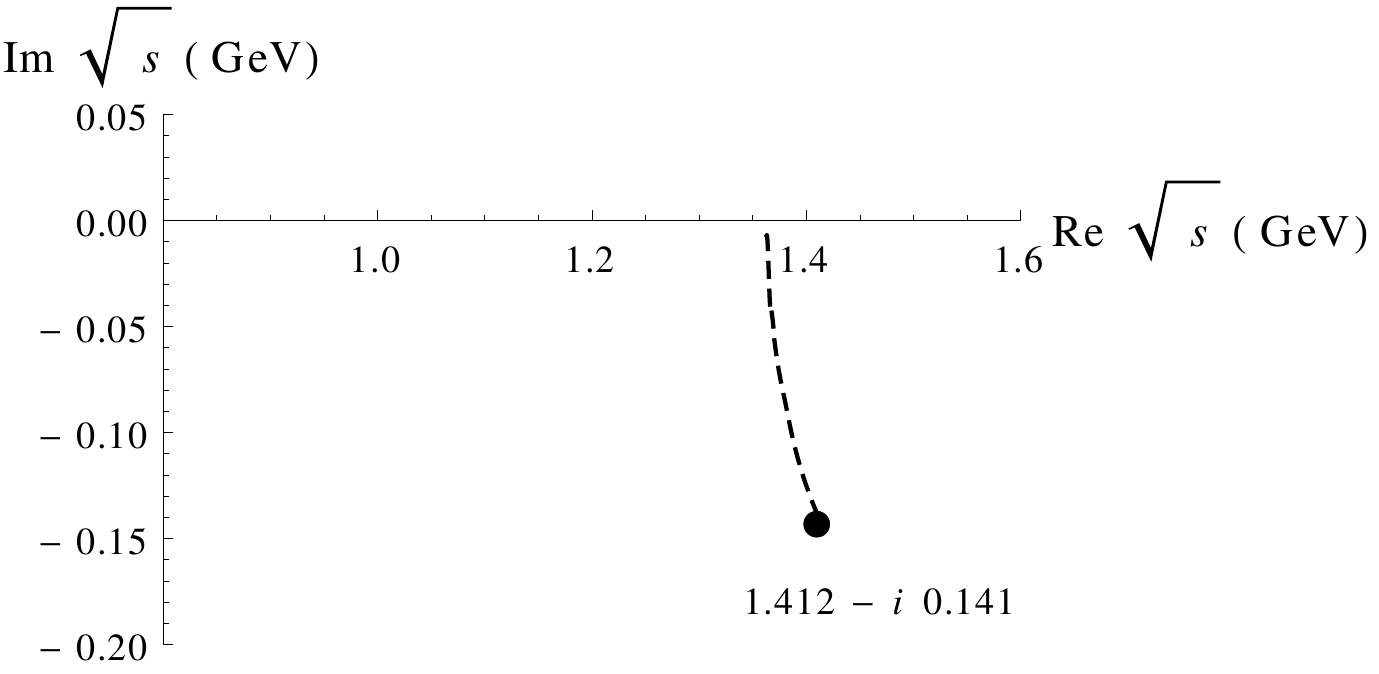}}};
\node (3dplot085) at (L085trajectory) [yshift=-1.1cm] [xshift=-2.0cm] {\fbox{\includegraphics[scale=0.29]{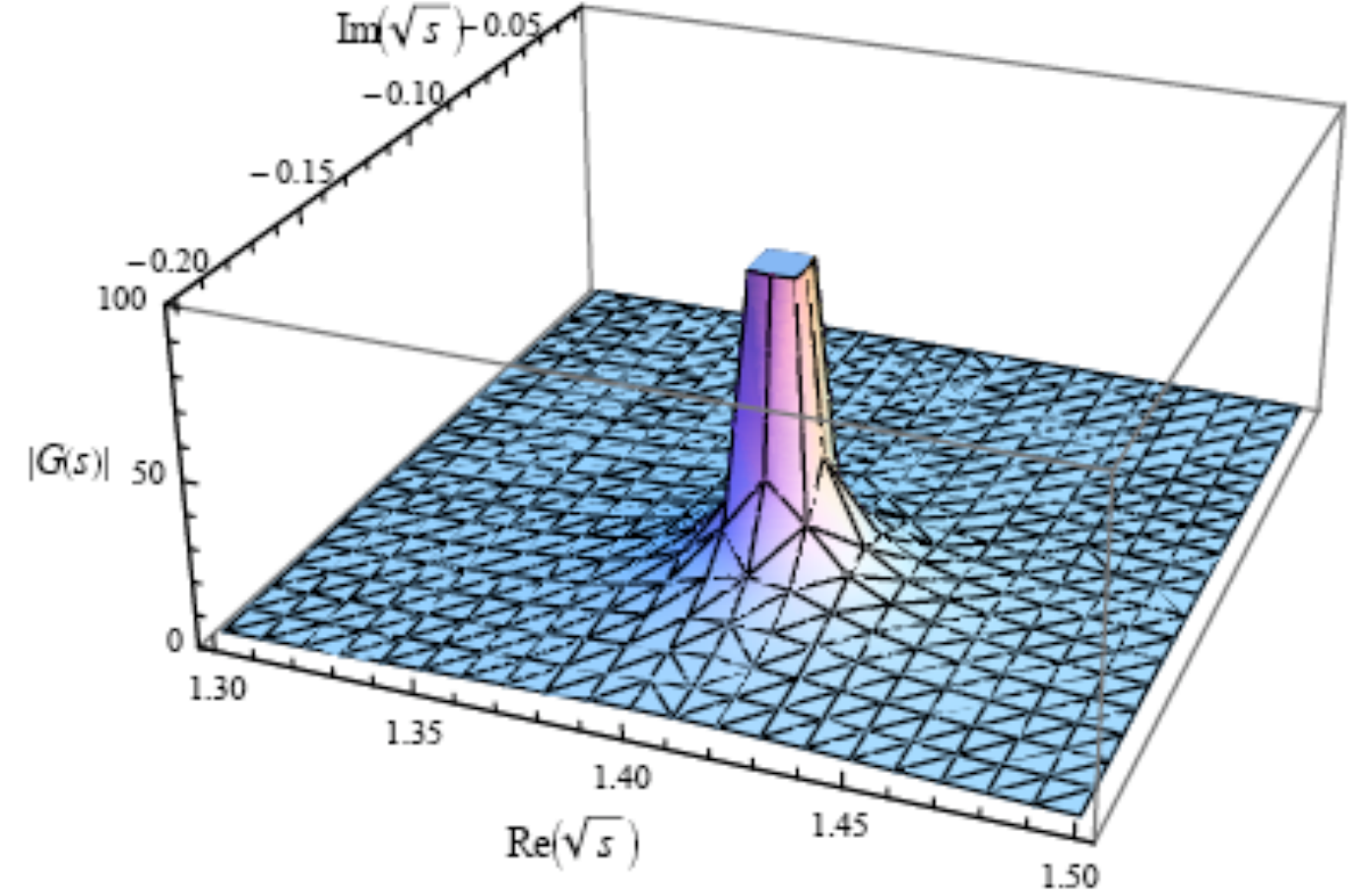}}};
\end{tikzpicture}
}
\caption{Position of the propagator pole for the $a_{0}(1450)$ in the case of $\Lambda=0.85$ GeV. The pole is located at $\sqrt{s}=(1.412-i0.141)$ GeV. The solid line indicates the pole trajectory for $\mathcal{L}_{\text{int}}\rightarrow g\mathcal{L}_{\text{int}}$ with $g=0...1$, where the pole moves down in the complex plane for increasing $g$.}
\label{figure_trajectory}
\end{figure}

We show in Fig. \ref{figure_trajectory} the position of the complex propagator pole on the sheet nearest to the physical region. The pole has coordinates $\sqrt{s}=(1.412-i0.141)$ GeV, giving a decay width of $\Gamma=282$ MeV (in good agreement with both the tree-level result from our model and the experiment) and a mass of $1.412$ GeV. The dashed line is the pole trajectory as function of $g$ with $\mathcal{L}_{\text{int}}\rightarrow g\mathcal{L}_{\text{int}}$ and $g=0...1$. While the bare mass $M_{0}$ coming from the eLSM differs from the experimental value by at least $\sim50$ MeV, the pole mass lies within the experimental error. Thus, the inclusion of loops represents an improvement of the tree-level results. However, all in all, the loop contributions have just a minor influence on the tree-level values of Ref. \cite{eLSM2}: this is important because it confirms that the fit of Ref. \cite{eLSM2} is robust. In the future, one should perform this check for all other broad states entering in the eLSM.

Another interesting observation is the fact that we do \emph{not} find a companion pole of $a_{0}(1450)$: the resonance $a_{0}(980)$ does not emerge for the values of the parameters determined in\ Ref. \cite{eLSM2}. This result is robust upon variations of the parameters. As a possible outlook for future work one should try to include the $a_{0}(980)$ as a tetraquark state into the eLSM \cite{tqmix} and/or perform a full scattering analysis so as to investigate the emergence of this resonance in more detail.

In conclusion, we have studied the popagator pole(s) of the isovector state $a_{0}(1450)$ as it is determined by the eLSM. It turns out that $(i)$ the obtained pole coordinates are in good agreement with the experiment and the tree-level results from the model, $(ii)$ nevertheless we find no additional pole that could be assigned to the corresponding resonance below $1$ GeV.

\bigskip

\textbf{Acknowledgments:} The authors thank D. H. Rischke, J.\ Wambach and G. Pagliara for useful discussions, and HIC for FAIR and HGS-HIRe for financial support.

\end{document}